\documentclass{article}
\usepackage{timet,color,graphicx}
\usepackage[urlcolor=blue,citecolor=black,linkcolor=black]{hyperref}
\usepackage{natbib}
\usepackage{geometry}
\begin{document}

\title{A tale of two faults: Statistical reconstruction of the 1820 Flores Sea earthquake using tsunami observations alone}
%\author[T. Paskett, J. P. Whitehead, R. A. Harris, C. Ashcraft, J. A. Krometis  et. al.]
  \author{T. Paskett, J. P. Whitehead, R.A. Harris, C. Ashcraft, J. A. Krometis,\and I. Sorensen, and R. Wonnacott}%\thanks{$^1$ Rincon Research Corporation, Tucson, AZ, USA \\ $^2$ Mathematics Department, Brigham Young University, Provo, UT, USA\\ $^3$ Geology Department, Brigham Young University, Provo, UT, USA\\ $^4$ United States Geological Survey, Menlo Park, CA, USA\\ $^5$ National Security Institute, Virginia Tech, Blacksburg, Virginia, USA\\ $^6$ Mathematics Department, Virginia Tech, Blacksburg, Virginia, USA\\ $^7$Mechanical Engineering, Brigham Young University, Provo, UT, USA}
%\date{Received 1998 December 18; in original form 1998 November 22}
%\volume{200}
%\pubyear{2023}

%\def\LaTeX{L\kern-.36em\raise.3ex\hbox{{\small A}}\kern-.15em
%    T\kern-.1667em\lower.7ex\hbox{E}\kern-.125emX}
%\def\LATeX{L\kern-.36em\raise.3ex\hbox{{\Large A}}\kern-.15em
%    T\kern-.1667em\lower.7ex\hbox{E}\kern-.125emX}
% Authors with AMS fonts and mssymb.tex can comment out the following
% line to get the correct symbol for Geophysical Journal International.
\let\leqslant=\leq

\newcommand{\jak}[1]{\footnote{\textcolor{blue}{ {\bf jak:}  #1 }}}

\newtheorem{theorem}{Theorem}[section]

%\label{firstpage}

\maketitle

\begin{abstract}
Using a Bayesian approach we compare anecdotal tsunami runup observations from the 29 December 1820 Flores Sea earthquake with close to 200,000 tsunami simulations to determine the most probable earthquake parameters causing the tsunami. Using a dual hypothesis of the source earthquake either originating from the Flores Thrust or the Walanae/Selayar Fault, we found that neither source perfectly matches the observational data, particularly while satisfying seismic constraints of the region. However, there is clear quantitative evidence that a major earthquake on the Walanae/Selayar Fault  more closely aligns with historical records of the tsunami, and earthquake shaking. The simulated data available from this study alludes to the potential for a different source in the region or the occurrence of an earthquake near where both faults potentially merge and simultaneously rupture similar to the 2016 Kaikoura, New Zealand event.
\end{abstract}

%\begin{keywords}
 %historical seismicity, tsunami, historical earthquake, Bayesian inversion
%\end{keywords}

\section{Introduction}
A thorough understanding of seismic history in tectonically active regions is necessary to determine the risk of future seismic hazards.  The challenge of this need is that faults have seismic events at time scales that stretch back well beyond only a century of instrumental records.  It is for this reason that there has been a focused effort to quantify past seismic events even though some observations may be unreliable \cite{newcomb1987seismic,sieh2008earthquake,meltzner2010coral,meltzner2012persistent,meltzner2015time,jankaew2008medieval,monecke20081,bondevik2008earth,bryant2007cosmogenic,grimes2006mapping,reid2016two,barkan2010tsunami,tanioka1996fault,nanayama2003unusually,LiuHarris2014,harris2016waves,fisherharris2016,GrNgCuCi2018,martin2019reassessment,ringer2021methodological}.  A significant concern with the reconstruction of these historical events is the inherent uncertainty that is unavoidably tied to the nature of the observations themselves.  Following the work of \cite{ringer2021methodological,krometis2021embracing} we apply a Bayesian framework to the task of quantitatively estimating the size and location of the Flores Sea Earthquake from 1820 that resulted in a devastating tsunami that was witnessed in four places throughout the Flores Sea region (Fig. 1).

As shown in \cite{ringer2021methodological,krometis2021embracing} the Bayesian approach provides a statistically justified method to generate several thousands of tsunami simulations, to determine the most probable source of the observed tsunami.  Not only does this approach provide a phenomenological approach toward sampling the earthquake parameter space, but it also automatically yields estimates on the uncertainty in those estimates as demonstrated below.  In the language of statistical inference, we are able to construct a posterior distribution on the set of earthquake parameters that best yields the observed tsunami characteristics.  This posterior distribution provides far more information than simply specifying a single earthquake that best fits the observational data, but is actually a probability distribution on all potentially valid parameters, thus specifying correlations between the different parameters of the earthquake. In addition the resultant simulated data provides a quantitative probabilistic assessment for the danger posed by a repeated tsunamigenic event of the same magnitude.

The focus of this article is on the December 29, 1820 earthquake that rocked SW Sulawesi leading to a devastating tsunami. Historical records \cite{wichmann1918earthquakes,wichmann1922earthquakes} document that the tsunami destroyed much of the settlement near Bulukumba on the SW arm of Sulawesi, and severely damaged the port city of Bima, Sumbawa over 300 kilometers away, as well as causing some tidal disturbance as far away as Sumenep on Madura Island off the NE coast of Java (Fig. 1).  For observations of this event, we rely on translations of the Wichmann catalog \cite{wichmann1918earthquakes,wichmann1922earthquakes,harris2016waves}, which details earthquakes and tsunamis of the Indonesian archipelago for parts of the 17th, 18th, and 19th centuries.  This particular event is of significant interest seismically as there are two potential sources of the earthquake: the Flores back-arc thrust (a hypothesis that is investigated in \cite{GrNgCuCi2018} for shaking observations of this event), and the more recently quiescent Walanae/Selayar Fault that parallels Selayar Island.  The impacts of such a major earthquake at either location on modern society would be devastating.  However, it is critically important to determine which of these faults was the source of the 1820 event in order to gauge the potential for future seismic hazards, particularly since there is evidence of Quaternary deformation of the Selayar Island region, but no significant instrumental earthquakes.  After constructing two posterior distributions, one for each potential fault source we quantitatively demonstrate that the Walanae/Selayar fault statistically is a far better fit to the observational data although it does not match the data perfectly.

The rest of this article proceeds as follows: The next section briefly reviews the Bayesian approach, and discusses the construction of prior distributions for the earthquake parameters i.e. models of the two disparate faults under consideration.  Section \ref{sec:likelihood} discusses the formation of a likelihood model that includes using the tsunami propagation model Geoclaw and the construction of the observational probability distributions.  Section \ref{sec:results} presents the results of the 200,000+ tsunami simulations including the use of a binary classification scheme to quantitatively determine that the Walanae/Selayar Fault is 90\% more likely to yield a match with modeled observations than the Flores thrust. Finally Section \ref{sec:conclusions} concludes with a brief discussion and explanation of the hypothesis of a multi-fault rupture and/or presence of an underwater landslide near Bulukumba.

\section{Bayesian Inverse Problems and  Construction of the Prior Distribution}
For the purposes of the current discussion, we briefly recall the basis for Bayes' Theorem and the use of Markov Chain Monte Carlo (MCMC) in identifying the earthquake parameters most likely associated with the 1820 event.  Rather than reviewing all of the details, we will provide a succinct summary and focus on those aspects of the inverse problem particular to this event.  A more detailed description of the approach taken here is provided in \cite{krometis2021embracing}, and more generally in \cite{gelman2014bayesian,kaipio2005statistical}.  We do, however focus on the application of Bayes' Theorem to the problem at hand, determining a reasonable probability distribution on parameters meant to model an earthquake given statistical observations of the resultant tsunami wave height and arrival time at different locations.

\subsection{Earthquake parameterization}
For earthquake induced tsunamis, we will consider earthquakes parameterized by the Okada model \cite{okada1985surface,okada1992internal} which is dictated by a set of 9 model parameters in three distinct categories:
\begin{enumerate}
    \item Magnitude (Mw):
    \begin{itemize}
        \item length $l$: the horizontal length of a rectangular rupture zone (typically measured in kilometers).
        \item width $w$: the width  of the same rectangular rupture zone (typically measured in kilometers).
        \item slip $s$: the amount of movement the rectangular rupture zone sustained during the seismic event (typically measured in meters).  Our model will assume a uniform slip distribution throughout the entire rectangular region.
    \end{itemize}
    The magnitude of the event can roughly be calculated as the logarithm of the product of these three variables.  We will specifically use the rectangular Okada model so that all ruptures are assumed to be adequately described by a series of connecting rectangles.
    \item Location:
    \begin{itemize}
        \item latitude $lat$: latitude coordinate of the earthquake centroid.
        \item longitude $lon$: longitude coordinate of the earthquake centroid.
        \item depth $d$: depth below the surface of the earth at which the centroid of the rupture occurs (typically measured in kilometers).
    \end{itemize}
    We assume that the fault ruptures instantaneously so that the epicenter and centroid are identical.  Further parameterization of a time-dependent, variable slip rupture is possible with the Okada model but we do not anticipate that our data is sufficiently robust to infer details for such a model.
    \item Orientation/geometry:
    \begin{itemize}
        \item strike $\alpha$: orientation of the fault measured clockwise in degrees from north.
        \item dip $\beta$: angle of inclination of the fault  from horizontal.
        \item rake $\gamma$: slip angle in degrees that the upper block of a fault (Hangingwall) moves relative to the strike angle, i.e. a rake of  $90^\circ$ corresponds to hanging wall slip up the fault parallel to the dip direction, which is a thrust fault.
    \end{itemize}
\end{enumerate}

\subsection{Bayesian inversion and MCMC}
Referring to all of these model parameters as the vector $\tilde{x} = (l,w,s,lat,lon,d,\alpha,\beta,\gamma)^T$, our goal is to determine a distribution on these nine parameters that best describes the 1820 earthquake by matching the historical record and our understanding of earthquake structure most closely.  Hence we seek to identify, or at least approximate the conditional probability distribution
$\pi\left(\tilde{x} | \mathcal{O}\right)$, where $\mathcal{O}$ are the historical observations that we have gleaned from the Wichmann catalog.  In other words we are going to approximate the probability of a specific set of Okada earthquake parameters, given the observations from the historical record.

The natural way to compute $\pi(\tilde{x}|\mathcal{O})$ is to apply Bayes' Theorem which states that this \emph{posterior} probability is proportional to the product of a \emph{prior} $\pi(\tilde{x})$ and \emph{likelihood} $L(\mathcal{O}|\tilde{x})$, i.e.
\begin{equation}\label{eq:Bayes_post}
    \pi(\tilde{x}|\mathcal{O}) \propto \pi(\tilde{x}) L(\mathcal{O}|\tilde{x}).
\end{equation}
The prior $\pi(\tilde{x})$ is a distribution that represents the \textit{a priori} expert knowledge of the potential distribution of earthquake parameters before examining the observational data, and $L(\mathcal{O}|\tilde{x})$ represents the likelihood of the historical observations occurring given a specific set of earthquake parameters $\tilde{x}$.  Specifying the prior and likelihood will then fully describe the desired posterior distribution.

The discussion above details the computation of the relative posterior probability for a particular set of parameters.  Approximating the full posterior distribution is a much more difficult task as~(\ref{eq:Bayes_post}) is only a relative proportionality i.e. the normalization of the full distribution is not available.  To adequately approximate the full distribution we use Markov Chain Monte Carlo \cite{gelman2014bayesian,kaipio2005statistical} which generates a Markov chain whose stationary distribution converges to the desired posterior.  For the computations performed here, we have utilized a random walk proposal kernel that takes randomized steps in parameter space between each proposed earthquake.  For example, suppose that we are currently considering earthquake parameters $\tilde{x}_k$, and have computed the prior probability $\pi(\tilde{x}_k)$, and after passing these parameters through our forward model (discussed below), also computed the likelihood $L(\mathcal{O}|\tilde{x}_k)$.
\begin{enumerate}
    \item A new set of earthquake parameters $\tilde{y} = \tilde{x}_k + \eta$ (referred to as the proposal) is proposed where $\eta$ is a random variable with a prescribed covariance matrix (chosen to yield the optimal mixing and convergence of the Markov Chain).
    \item The prior and likelihood of $\tilde{y}$ are computed as well.
    \item The proposal $\tilde{y}$ is \emph{accepted} based on the relative probability:
    \begin{equation}
       \alpha = \min\left( \frac{\pi(\tilde{y})L(\mathcal{O}|\tilde{y})}{\pi(\tilde{x}_k)L(\mathcal{O}|\tilde{x}_k)},1\right),
    \end{equation}
    i.e. we accept the proposal if it has a relatively higher probability than the current sample $\tilde{x}_k$, but may also accept the proposal (with lower probability) even if the posterior probability is less.
    \item If the proposal is accepted then $\tilde{x}_{k+1}=\tilde{y}$ and otherwise $\tilde{x}_{k+1}=\tilde{x}_k$.
\end{enumerate}

The sampling procedure introduced above is a bit too simplistic for the situation at hand.  As noted in \cite{ringer2021methodological,krometis2021embracing}, the standard Okada model parameters $\tilde{x}$ are correlated with each other, and hence it is not practical to search over each of these parameters separately.  Instead, we note that the geometry and depth of the fault explicitly depend on the latitude/longitude location of the epicenter, and the 3 magnitude parameters are highly correlated with respect to the magnitude itself.  We address these issues by introducing \emph{sample} parameters $x$ that we search over, from which the \emph{model} parameters $\tilde{x}$ can be computed, i.e. $\tilde{x} = f(x)$ for some map $f$.

The sample parameters are derived from the two different observations noted above:
\begin{itemize}
    \item The length, width, and slip are used to compute the magnitude and are highly correlated, i.e. the aspect ratio of an earthquake rupture zone follows a relatively deterministic log-linear relationship.  As described in detail in \cite{ringer2021methodological} this allows us to sample instead from magnitude $M$, and both of $\Delta \log l$ and $\Delta \log w$ which are deviations from the log-linear relationship between magnitude, length, and width which is identified from a log-linear fit to data from the past 70 years of earthquakes.
    \item The depth, strike, rake, and dip can be well approximated as functions of the latitude and longitude given previous fault plane solutions for more recent earthquakes along each of the two faults in question.  However as we don't completely believe/trust the existing modern data nor the model selected to represent it (described below), we also introduce the offset sample parameters: depth\_offset $\Delta d$, strike\_offset $\Delta \alpha$, dip\_offset $\Delta \beta$, and rake\_offset $\Delta \gamma$ which represent adjustments to the modeled geometry of the fault. 
\end{itemize}

\begin{figure}
    \centering
    \includegraphics[width=.5\textwidth]{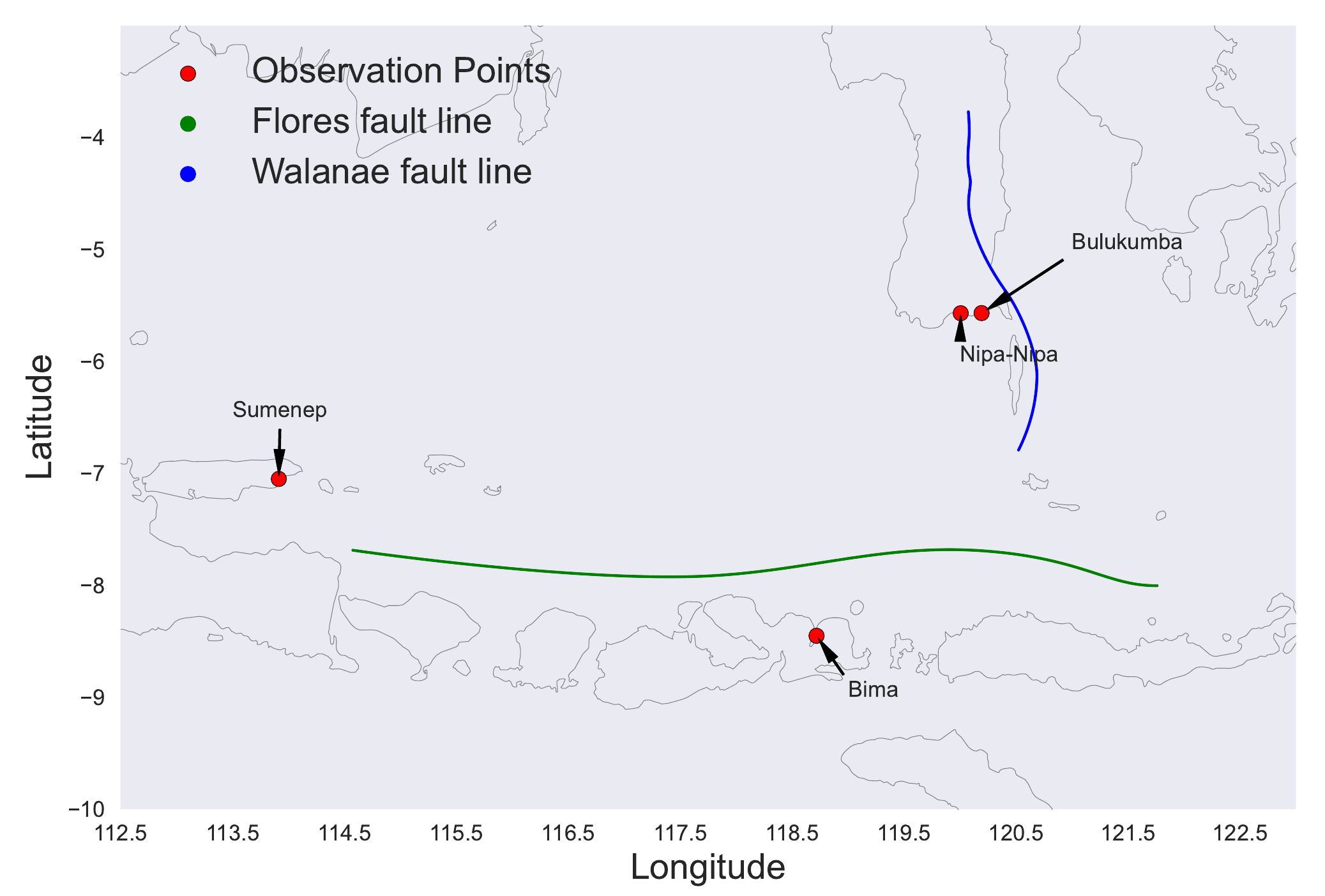}
    \caption{A depiction of the latitude longitude placement of the Walanae/Selayar Fault (blue, striking N-S) and Flores Thrust (gtriking E-W).  The red dots are the locations of tsunami observations with Sumenep to the far west, Belukumba and Nip-Nipa to the NE on the SW arm of Sulawesi, and Bima in Sumbawa (south of the Flores thrust).}
    \label{fig:faults_obs_pts}
\end{figure}

\subsection{Modeling the faults}
To develop the prior distribution for each fault we create a simplified model based on existing fault-plane solutions.  This leads to two very different prior distributions as there is a substantial amount of data to constrain the Flores Thrust, but very little to constrain the geometry and location of the Walanae/Selayar Fault.

\subsubsection{Modeling and sampling from the Flores Thrust}
The Flores Thrust forms in the backarc region of the eastern Sunda and Banda volcanic arcs due to distribution of strain away from the arc-continent collision occurring in the region \cite{hamilton1979tectonics,silver1983back,harris2011nature}. The fault is inclined to the south and moves the volcanic arc northward over the Flores Sea ocean basin. This motion is driven by the  high frictional resistance to subduction of the Australian continent beneath the volcanic arc. The amount of convergence between the Australian and Asian Plates that is partitioned to the Flores thrust increases eastward from 21-58 percent \cite{nugroho2009plate}. The two largest recorded earthquakes on the fault were in 1992 (Mw 7.8) and 2004 (Mw 7.5). Both of these earthquackes generated tsunamis, but neither impacted the areas inundated by the 1820 event. The USGS earthquake catalog lists over one hundred other recorded earthquakes along the Flores thrust, however a large number of these do not have full fault-plane solutions, or are missing some component of the needed fault geometry parameters.  After filtering these data to restrict earthquakes exceeding $5.0$ Mw, and with the following parameters defined:
\begin{enumerate}
    \item latitude-longitude of the hypocenter
    \item depth
    \item dip
    \item strike
\end{enumerate}
we were left with 94 seismic events in the instrumental record.  These fault plane solutions formed the basis for our prior distribution on Flores thrust fault geometry.

\begin{figure}
    \begin{center}
    \includegraphics[trim = 0mm 0mm 0mm 0mm, width=.5\textwidth]{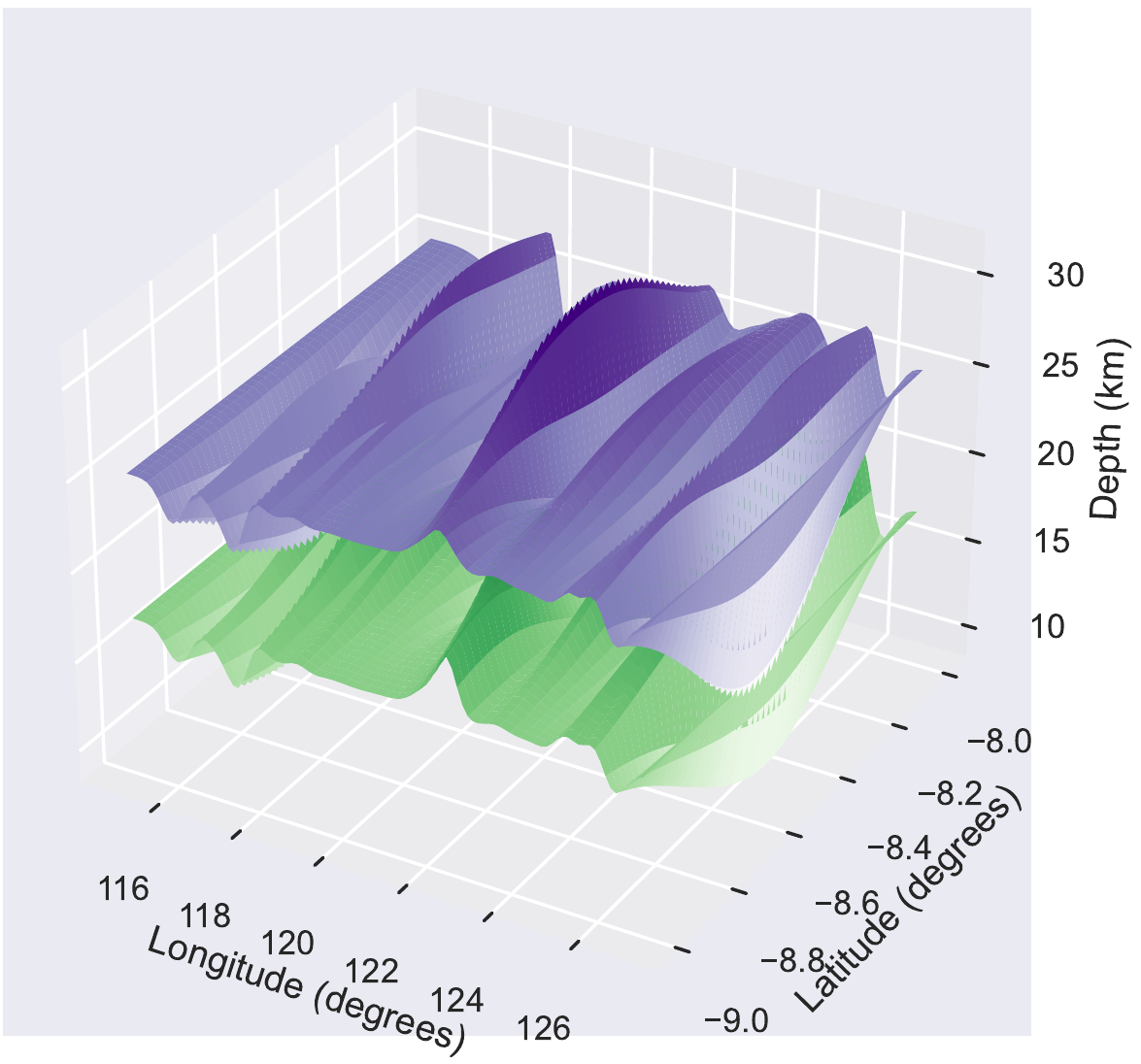}
    \end{center}
    \caption{The two surfaces defining one standard deviation away from the mean fit for the depth of the Flores thrust.  The Gaussian process for the depth defines the most probable depth as the region between these two surfaces.  Note the significant difference in scales between the two axes: this represents a change of only one degree in latitude but 10 degrees in longitude.  The difference in scales explains the apparent `ridged' behavior of the two surfaces along the longitudinal direction.}
    \label{fig:GP_depth}
\end{figure}

Due to the noisy and inherently irregular nature of this collected earthquake source data, we first created a multidimensional Gaussian process \cite{williams2006gaussian} to represent/model the Flores thrust.  This was done by considering the depth, dip, rake, and strike as independent functions of the hypocenter latitude and longitude of each instrumentally recorded event, and developing a statistical Gaussian process fit using a radial basis function (rbf) kernel with variance $0.75$ and a normalized noise level in the data itself of $1.0$ (see Algorithm 3.2 of \cite{williams2006gaussian} for details).  The benefit of using a Gaussian process rather than a standard regression technique is that under the assumed hyperparameters (variance of the kernel etc.) then the uncertainty is built into the regression.  This is demonstrated in Figure \ref{fig:GP_depth} which depicts two depth surfaces that correspond to depths that are one standard deviation away from the mean predicted depth, i.e. roughly speaking we anticipate that approximately two thirds of the earthquakes on the Flores thrust will be contained between these two surfaces.  Similar processes are constructed for the dip, rake and strike of the fault as well.

All parameters of the Flores thrust are modeled by these four Gaussian processes treated independently.  The prior distribution is then selected to match this model.  As discussed in \cite{ringer2021methodological} we develop a prior distribution on the latitude-longitude of the hypocenter by enforcing a distribution on the mean depth computed for our fault model.  We use a Gaussian distribution on depth with mean $30$ km and a standard deviation of $5$ km with a truncation on the interval $[2.5,50]$ km.  Hence each latitude-longitude coordinate is mapped through the model and the mean depth is then used to calculate a prior probability.  The mean dip, rake, and strike 
are then computed from the Gaussian process model and we sample over the novel offset parameters: depth\_offset, dip\_offset, rake\_offset, and strike\_offset which allow for perturbations from the mean statistical model.  To compute the final Okada earthquake parameters, we take the computed mean depth, dip, rake, and strike and then add the standard deviation of the Gaussian process at that point multiplied by the corresponding offset parameter.

In summary, in addition to the three parameters prescribed for the magnitude of the earthquake we introduce the following sample parameters:
latitude, longitude, $\Delta d,~\Delta \alpha,~\Delta\beta,$ and $\Delta\gamma$.  These are mapped through the Gaussian process fault model and then the offset parameters are used to produce the Okada earthquake parameters: latitude, longitude, depth, dip, rake, and strike.

\subsubsection{Walanae/Selayar Fault}
Earthquakes are recorded for most of the Walanae/Selayar Fault (17 events $>3.0$ Mw)  including 3 quakes of Mw 5.0-5.9 since 1993 \cite{jaya2020paleoseismic}.   However, the section of the fault south of Bulukumba (Belokumba), known as the Selayar Fault, which causes uplift of Selayer Island, is under-slipped with 5-10 mm/a of convergence to the ENE. This fault, which causes uplift of Quaternary coral terraces on Selayar Island, currently may be in a phase of interseismic elastic strain accumulation, but is capable of generating a tsunami,\cite{sarsito2019walanae,simons2007decade} \cite{cipta2017probabilistic}. Lack of instrumentally recorded earthquakes on the Selayar Fault hinders efforts to properly fit a Gaussian process to model the fault.  Limited detail and constraint on the existing data lead us to make a simpler hypothesis for the fault parameters.  We modeled the Walanae/Selayar fault as a plane following a default dip of $25^\circ$ i.e. for a given latitude longitude the depth of the fault is calculated assuming that the fault interface dips $25^\circ$. The fault strike is measured from different geographic points parallel to the fault and projected perpendicular to the fault line to points interior to the fault itself. We assume that the rake on the fault is centered at $80^\circ$ throughout as there is no data to constrain the rake any further.

To account for the uncertainty in this over-simplified model of the Walanae/Selayar fault, we also introduce and search over $\Delta d,~\Delta \alpha,~\Delta\beta$ and $\Delta \gamma$, thus allowing for some strike-slip motion which is evident on the Walanae section of the fault.  In contrast to the Flores thrust, the final Okada parameters are then obtained from simply adding the offset parameters to those computed from the planar model (the offsets in the Flores thrust are first multiplied by the corresponding standard deviation from the Gaussian process fit).  This leads to the final set of Okada parameters required by the forward model.

\section{Construction of a Likelihood function}\label{sec:likelihood}
\subsection{Observational probabilities}
As described above, we make use of extremely anecdotal observational accounts that present a high level of uncertainty. For instance the historical account records that in southern Sulawesi (see \cite{wichmann1918earthquakes,wichmann1922earthquakes}
\begin{quote}
...there was after a weak shock, vibrations becoming gradually more powerful, such that the flat of the commandant in Fort Bulekomba fluctuated to and fro. The six-pounders set up in bastion number 2 hopped from their mounting. After the 4-5 minute long quake, shots were believed to be heard in the west, coming from the sea. Barely had the sent envoy returned with the news that ships were nowhere to be seen, than did the sea, under a both whistling and thunder-like rumble, come in, formed as a 60-80 foot high wall, and flooded everything.
\end{quote}
This particular account is unique because it yields both an arrival time (after the main earthquake) of the initial wave and an approximate wave height.  This also clearly illustrates the anecdotal and uncertain nature of the observational data that we are using.  There is very little in the way of definitive measurements that can be used to pin down the exact nature of either the earthquake or the subsequent tsunami.  The hypothesis is that a combination of several such observations will be enough to adequately constrain some of the earthquake parameters to at least partially glean information on the causal earthquake.

We have chosen to focus on observations of the tsunami alone, as shaking intensity is notoriously a highly uncertain prediction \cite{abrahamson2016bc} particularly without extensive knowledge of VS30 at each observation site.  Precise measurements and careful study of the entire Flores Sea region may yield a set of Ground Motion Prediction Equations (GMPE) that fits the ground motion, but to date no such data is available (see \cite{GrNgCuCi2018} where such a study is carried out with a generic GMPE).  As we have a physics-based and rigorously validated \cite{berger2011geoclaw} forward model for tsunami propagation, we are more confident in inferring earthquake parameters from observations of the tsunami.  Although we do not make direct use of the shaking observations, the historical record of shaking intensity can be used to validate our results as discussed in Section \ref{sec:conclusions}.

As already described, the textual observation cited above illustrates the two types of observations that we make use of for the 1820 tsunami at different geographic locations:
\begin{itemize}
    \item Wave arrival time: The time it takes for the initial wave to reach a specific location.
    \item Maximum wave height: The maximal wave height at a specific location.
\end{itemize}
For the 1820 tsunami we identified 4 distinct geographic locations around the Flores Sea where the tsunami was observed (Fig. 1).  There are a few things worth pointing out about these observation locations before we consider the actual observations themselves.
\begin{enumerate}
    \item Bulukumba and Nipa-Nipa are both on the southern tip of SW Sulawesi and 20 km apart.  The historical record reports that the earthquake lasted 4.5 minutes that was followed by tsunami 18-24 m high at the Fort Bulukumba that inundated 300-400 m inland destroying villages around Nipa Nipa  and carrying ships off the coast into rice fields.  
    \item Sumenep is over 700 kilometers WSW of Bulukumba over a relatively shallow sea (much of the Flores Sea is less than $300m$ deep) so a wave that reaches both locations would dissipate a significant amount, and take a long time to propagate that far.
    \item Bima, on Sumbawa Island (the southern most observation location), is deep inside a narrow inlet that opens into a bay.  It is well known that inlets and bays \emph{can} amplify tsunamis, but the angle of incidence in such a case is critical to capture the effects accurately and capturing such an effect may require simulations at a higher resolution than the available bathymetry allows.
\end{enumerate}

With all of these considerations in mind, we define the observational probability distributions for each observation on a case-by-case basis, some of which are illustrated specifically here.
\begin{itemize}
    \item To begin, the account quoted above that refers to the wave height being 18-24 m near Bulukumba is likely an over-exaggeration than under so the observational probability distribution on wave height at Bulukumba is a normal distribution centered at $18m$ with a standard deviation of $5m$.
    \item Similarly the wave arrival time at Bulukumba is prescribed as a normal distribution centered at 15 minutes with a standard deviation of 10 minutes (truncated at a 0 minute or instantaneous arrival). This is based on the proximity of Fort Bulukumba to the coast.
    \item The observation at Bima is given by
    \begin{quote}
        Bima on Sumbawa. Violent quake of a good 2 minutes in duration, which was followed by a violent rumbling and then a flood wave that flung anchored ships far inland and over roof tops.
    \end{quote}
    As there is no time given we make use of the observation of wave height only.  Although flinging ``anchored ships far inland'' is very graphic, it’s not very quantitative. The fact the ships were anchored seems to indicate they were larger than say, just canoes or other small boats. This observation, and the fact that they were flung ``far'' inland and over roof tops, indicates a sizeable wave. We don’t think that waves smaller than 1 meter are plausible. So, for Bima’s wave height we chose a truncated Gaussian likelihood with mean 10 meters, standard deviation 4 meters, and a lower bound of 1 meter.
    \item The account from Nipa-Nipa has no estimate of wave height but only inundation, which leads to an observational distribution with an assigned mean of 3 meters. The tsunami striking Sumenep was observed without any detail so we select a truncated distribution centered around 1.5 meters (basically guaranteeing a wave of some sort is noticed at Sumenep).
    \item The final observation is the wave arrival time at Sumenep.  In this case the historical record indicates that the wave arrived at Sumenep 5 hours after the earthquake was felt in Bulukumba and Bima.  The issue with this particular observation is that Bima and Bulukumba are currently (and according to Dutch records was at the time) in a different time zone than Sumenep.  In particular, the record indicates that the earthquake was felt close to 10:00 hours, but the wave arrived in Sumenep at 15:00 hours.  The issue is that Sumenep is on the very eastern edge of its time zone, and at different times in the 1800s, was either in the same time zone as Bulukumba and Bima, or 30 minutes or 60 minutes off.  In addition to the concerns over the time zones which were not standardized in Indonesia until 1912 \cite{nguyen2015indonesia}, the definitive times of 10:00 and 15:00 hours are rather ambiguous (if the record had instead cited 10:12 and 15:27 for instance, then we would take more credence to the precise time interval).  All of this is to say that although this observation does indicate that the wave took a very long time to reach Sumenep, the exact timing of the wave's arrival is very clearly uncertain.
    
    From preliminary estimates of the wave speed across the Flores Sea (recall that in open water tsunamis travel very near the linear phase speed $\sqrt{gH}$ where $g$ is the gravitational constant, and $H$ is the water depth), we were unable to legitimately justify a wave originating from any location on either proposed fault and taking even close to 5 hours to reach Sumenep.  Hence to construct the observational probability distribution for the arrival time at Sumenep, we went with the hypothesis that Sumenep was in a different time zone than the other observation locations which would put the observed time interval at 4 hours rather than 5.  With this in mind, we selected a normal distribution with a mean of 240 minutes (4 hours) and a standard deviation of 45 minutes.
\end{itemize}
The final observational probability distributions are illustrated in Figure \ref{fig:Observations} as the continuous red curves.

\subsection{The forward model}
As discussed in more detail in \cite{ringer2021methodological} we make use of Geoclaw as the forward model which takes the required earthquake parameters as inputs, and applies the Okada model \cite{okada1985surface,okada1992internal} to generate an idealized seafloor deformation which is then used as an initial condition for the fully nonlinear shallow water equations.  Geoclaw has the capability of rendering both rectangular and triangular faults, but we only take advantage of the former.  Unlike the Banda Arc studied in \cite{ringer2021methodological} both the Flores Thrust and Walanae/Selayar Fault are fairly geographically linear and hence are easily modeled by a small number of rectangular faults.  In particular we use three rectangular faults to model the full rupture zone of each fault.

The Okada rectangular rupture regions are identified via the following process which is a simplification of that employed for the 1852 event in \cite{ringer2021methodological}.
\begin{enumerate}
    \item The latitude-longitude centroid location is identified via the random walk Monte Carlo step, and the total width and length of the rupture are computed from the sampled magnitude and $\Delta \log l$ and $\Delta \log w$ as described above.
    \item The length is split into 3 and the rupture is specified as three different rectangular regions, each with the same width.  The centroid of each of these rectangles is identified along a line of equal depth according to the model specified for each fault (the Gaussian process for the Flores Thrust etc.) and the orientation is parallel to the modeled fault.
    \item The Okada model is employed for each of the three sub-rectangles for a simultaneous, instantaneous rupture.
\end{enumerate}

Following the formation of the seafloor deformation from the 3-rectangular rupture via the Okada model, Geoclaw uses a finite volume formulation \cite{berger2011geoclaw} with a dynamically adaptive spatial mesh to simulate the propagation of the resultant tsunami via the nonlinear shallow water equations.  We leave most parameters in Geoclaw as their default values including bottom drag and friction coefficients, and carefully tune the adaptive mesh as described below.

The forward propagation of a tsunami wave critically depends on accurately resolving the bathymetry (underwater topography), which is a difficult and pressing issue for all tsunami simulations and studies.  For bathymetry we primarily relied on the 1-arcminute etopo
datasets available from the open access NOAA database (\url{https://www.ngdc.noaa.gov/mgg/global/global.html}), and for the coastline near each
observational point we utilize higher resolution Digital Elevation
Models (DEM) from the Consortium for Spatial Information (CGIAR-CSI, \url{http://srtm.csi.cgiar.org/srtmdata/}).  These higher resolution
topographical files yield a 3-arcsecond resolution on land, but give
no additional information on the sub-surface bathymetry.  In addition we also took advantage of detailed sounding maps available from the Badan Nasional Penanggulangan Bencana (BNPB or Indonesian National Agency of Disaster Countermeasure, see \url{http://inarisk.bnpb.go.id}).  To convert these data into digitally accessible information, contours were taken from images exported from the website and then traced and interpolated in arcGIS to produce approximate depths in the same regions as the DEM files.  This approach provides a set of bathymetric files that are accurate to around 10-15 arcseconds near each observation location with a maximum possible resolution of 3 arcseconds.

We make use of six different levels of refinement, starting with a resolution of 6 arcminutes in the open ocean going down to 3 arcseconds (the maximum resolution allowed by our bathymetric data) around those parts of the wave that will impact the observation locations directly (see \cite{ringer2021methodological} for a more thorough description of the same adaptive mesh).  The mesh refinement is activated whenever the solution of the linearized backward adjoint equation \cite{DavisLeVeque2016} exceeds a specified threshold at the same time that the forward solution does as well.  The linearized adjoint solution is computed on a global mesh of 15 arcseconds, initialized with an endpoint condition corresponding to pointwise Gaussian sea surface perturbations at each observation location so that the adjoint solution solved backward in time will identify when and where the forward tsunami will be that directly affects each of the observation locations.  This dictates where the mesh is refined.  The benefit of using the adjoint driven adaptive mesh is that because every one of our Monte Carlo samples uses the same observation locations, then we need only run the linearized adjoint solver one time (hence the global 15 arcsecond resolution, while expensive, is a one time cost), and save the corresponding output to be used with the forward runs.

In addition to the dynamically adaptive mesh, we include several statically refined regions at the highest (3 arcsecond) resolution.  Each of these regions is specified as a series of rectangular (Geoclaw requires specification of regions in rectangular latitude-longitude coordinates) sub-regions that encapsulate each observation location.  This is meant to ensure that the incoming wave is accurately captured as it approaches each observation location.
For instance, Bima in Sumbawa is located deep inside a bay that must be accurately captured in order to simulate the tsunami reaching Bima, and so we defined several statically defined regions that encapsulate the bay and surrounding coastline as much as possible without unnecessarily refining the grid on land at the same time. 

\begin{figure}
    \centering
    \includegraphics[width=.45\textwidth]{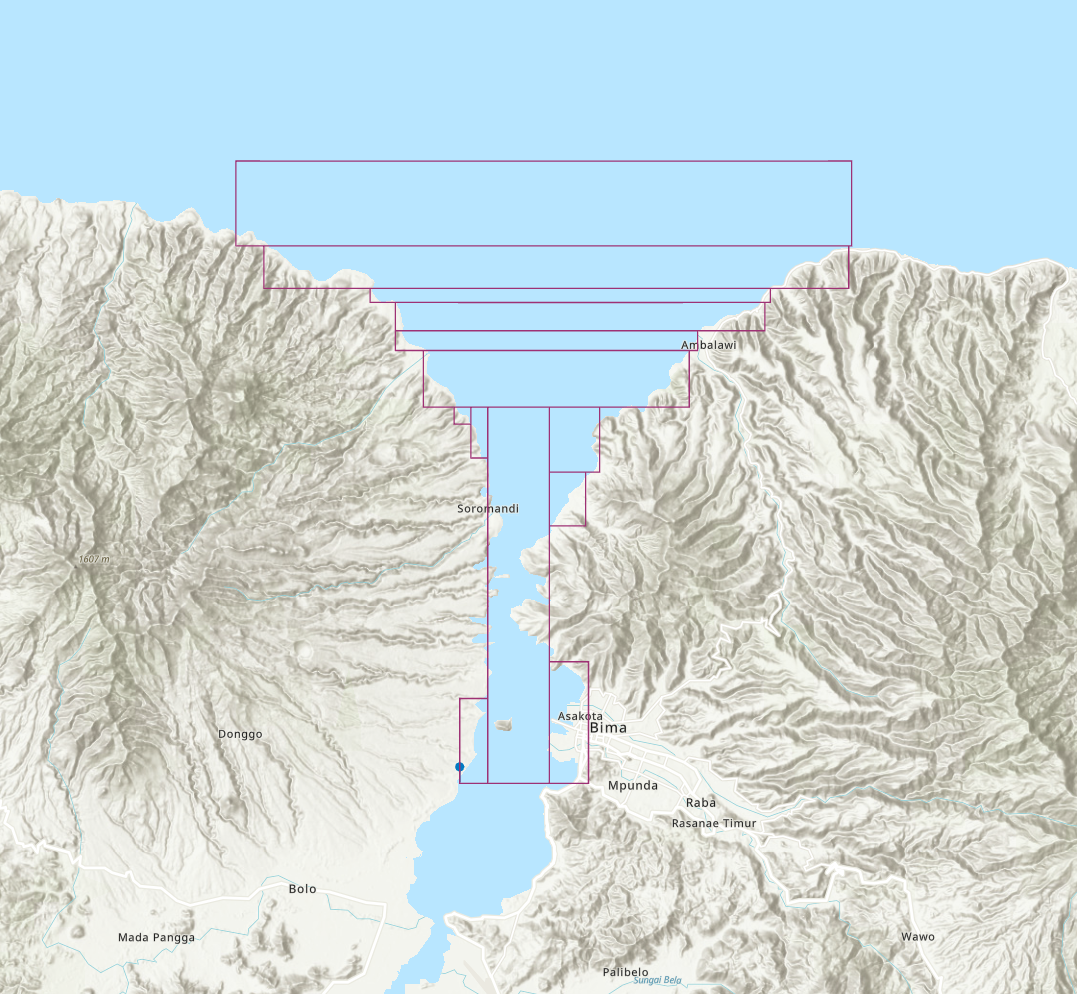}
    \caption{The rectangular regions where the resolution is fixed at the highest grid level near the port of Bima.  As Geoclaw requires each region to be specified as a rectangular region, we specified several sub-regions (shown as the red rectangles) that depict the regions of interest.  Similar highly resolved regions are defined for all of the other observation locations as well.}
    \label{fig:bima_resolve}
\end{figure}

We ran each tsunami simulation for at least 4 hours in physical time (we initially ran the tsunamis for 5 hours, but none of the waves required more than 4 hours to reach Sumenep, so we allowed the samples to run for 4 hours only to save compute time).  Running on 24 cores on a single node each of these simulations took approximately 10-12 minutes of wall-clock time, i.e. 240-288 minutes of compute time.  Wave heights and arrival times were extracted from the Geoclaw output using the previously developed tsunamibayes package \cite{zenodov1_1} and wrapped into the MCMC method to create the optimal sampling strategy.

%JPW
\section{Results}\label{sec:results}
\subsection{Statistical summary}
For each fault we initialized ten different chains with five unique latitude-longitude locations geographically spread across the entire fault and with two different magnitudes: $8.0$ and $8.5$ for a total of ten initial earthquakes.  After running each chain for two thousand samples a piece, we resampled all ten chains according to their final posterior probability and restarted each chain accordingly.  In this process most of the chains were eliminated, as most had still not achieved a finite log likelihood (most of the chains were unable to generate a noticeable tsunami wave that reached Sumenep).  After resampling, each chain was run for a minimum of 9,000 samples via  random walk MCMC.  In total, we simulated 104,970 tsunamis originating from the Walanae/Selayar fault and 127,690 originating from the Flores thrust.  This cost an estimated 110 years of total compute time spread over 24 cores at a time and twenty chains, for nearly 2.5 months of real time computational cost.

\begin{figure}
    \centering
    \includegraphics[width=.5\textwidth]{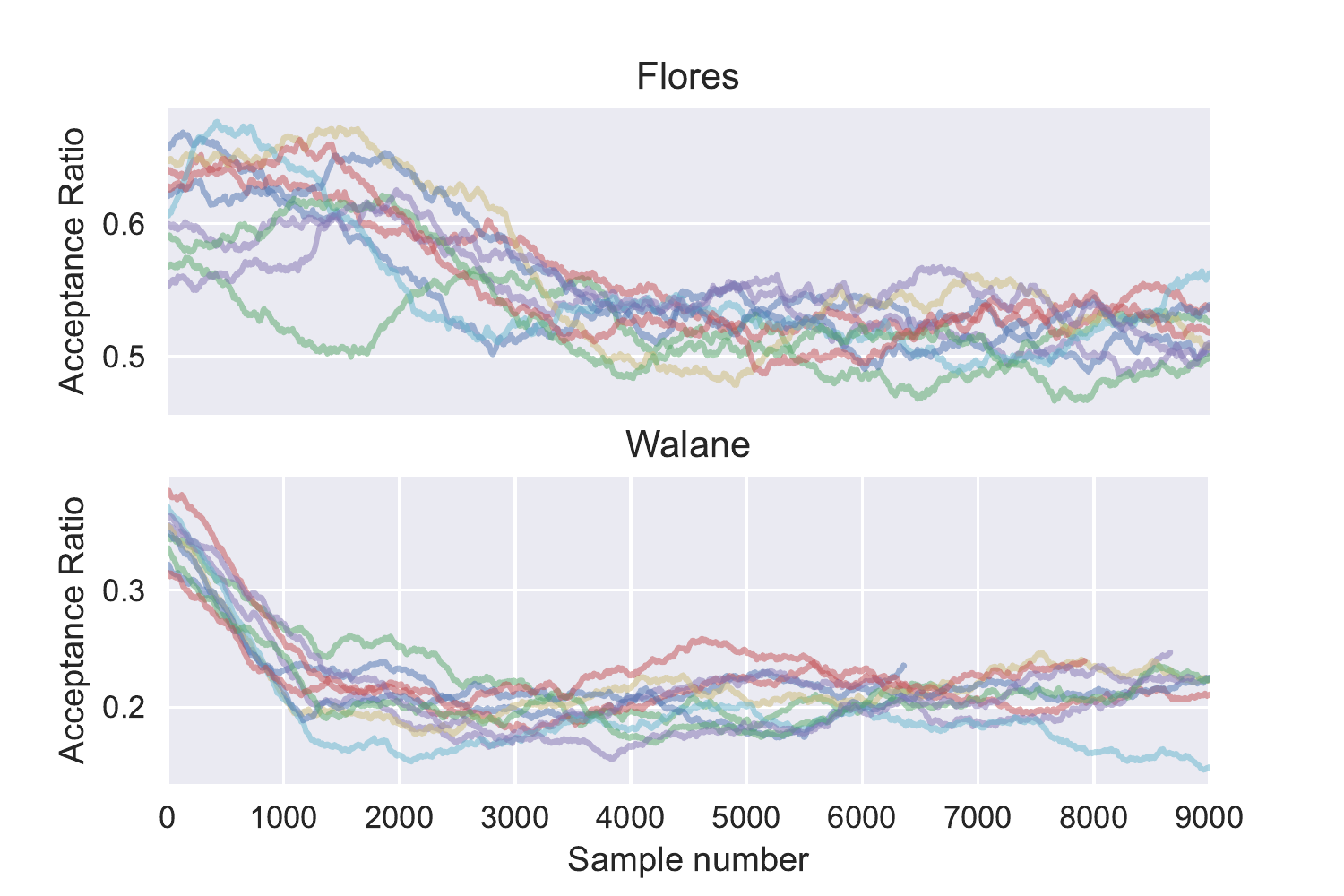}
    \caption{The acceptance ratio averaged over a 2000 sample interval for each fault's set of chains.}
    \label{fig:accept_ratio}
\end{figure}

The random walk step was initiated according to a diagonal covariance matrix with entries corresponding to the following for each sample parameter:
\begin{itemize}
    \item latitude (degrees): 0.086 (Flores), 0.05 (Walanae/Selayar)
    \item longitude (degrees): 0.11 (Flores), 0.04 (Walanae/Selayar)
    \item magnitude (Mw): 0.075 (Flores), 0.045 (Walanae/Selayar)
    \item $\Delta \log l$: 0.0132 (Flores), 0.012 (Walanae/Selayar)
    \item $\Delta \log w$: 0.0132 (Flores), 0.012 (Walanae/Selayar)
    \item $\Delta d$ (km): 0.525 (Flores), 0.55 (Walanae/Selayar)
    \item $\Delta \beta$ (degrees): 2.7 (Flores), 2.55 (Walanae/Selayar)
    \item $\Delta \gamma$ (degrees): 3.7 (Flores), 3.55 (Walanae/Selayar)
    \item $\Delta \alpha$ (degrees): 3.15 (Flores), 3.05 (Walanae/Selayar)
\end{itemize}
This covariance matrix was adjusted slightly (covariance values for dip offset $\Delta \beta$ and rake offset $\Delta \gamma$ as well as the longitude for Flores only were increased partway through the sampling) with the goal of getting close to a $0.25$ acceptance ratio for both sets of chains.  The averaged acceptance ratio for each different set of chains is depicted in Figure \ref{fig:accept_ratio}.  Note that the acceptance ratio for the Walanae/Selayar chains is slightly below the desired value, but the acceptance ratio for the Flores chains is still quite high, indicating that the sampling may be more aggressive on the Flores thrust then the covariance matrix described above.  Despite this high acceptance rate, all ten chains were mixing very nicely in all of the relevant variables.

\begin{figure}
    \centering
    \includegraphics[width=.5\textwidth]{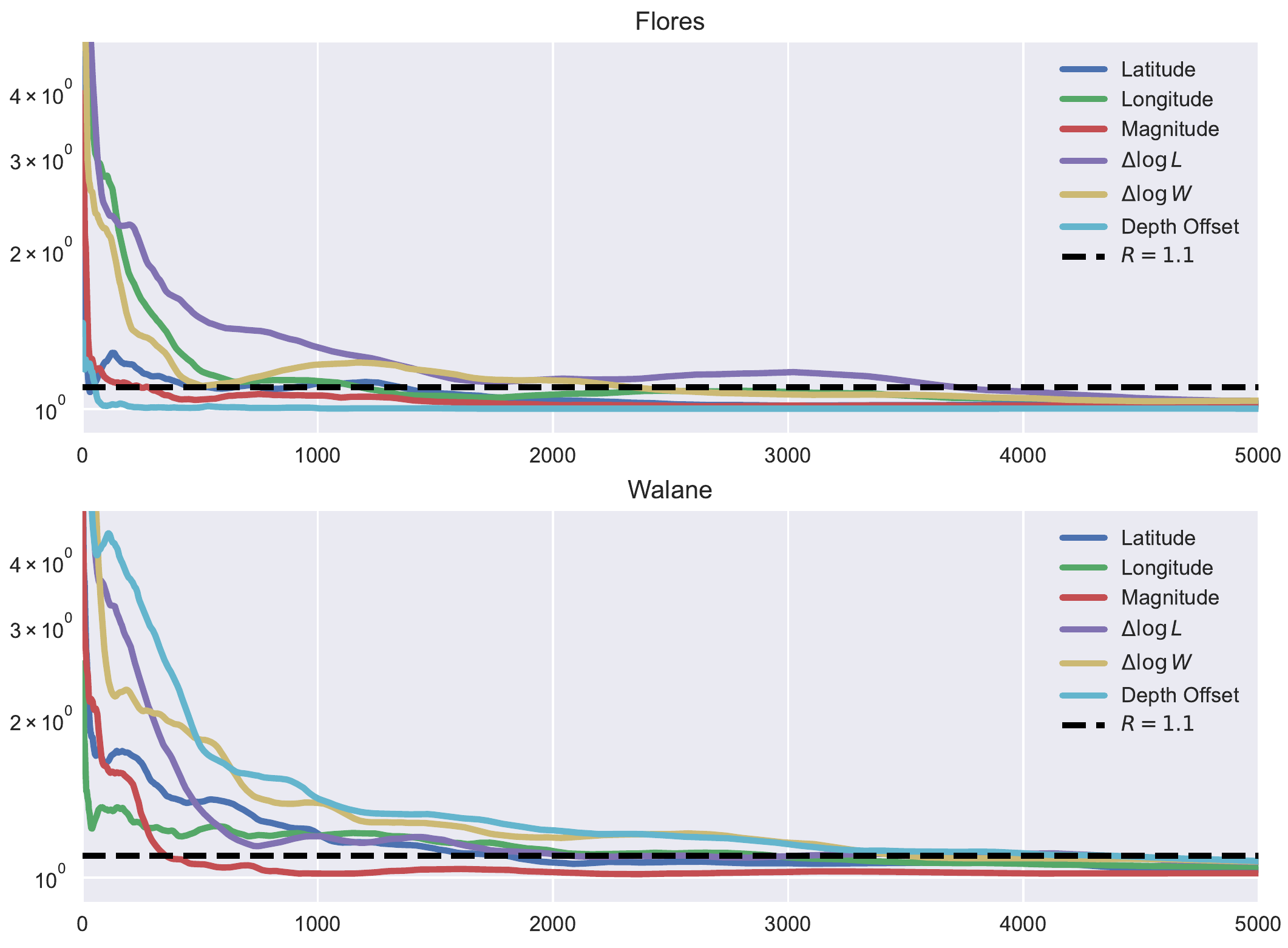}
    \caption{The Gelman-Rubin diagnostic displaying the relative inter-chain vs. intra-chain variance for each constructed posterior distribution.  The dashed horizontal line is at the value of $1.1$.  As described in \cite{gelman1992inference,gelman2014bayesian} the diagnostic should drop below this line to indicate appropriate mixing of the sampled posterior.  Note that this occurs for both posteriors before 5,000 samples are collected (for every observable computed here) so we do not show the rest of the chain's data for samples beyond 5,000.}
    \label{fig:gelman_rubin}
\end{figure}

To verify the inter-chain mixing and ensure that the approximated posterior distribution is adequately converged, we computed the Gelman-Rubin diagnostic \cite{gelman1992inference,gelman2014bayesian} for all of the parameters from the posterior distribution as shown in Figure \ref{fig:gelman_rubin}.  Note that the Flores posterior mixes at a slightly faster rate (the Gelman-Rubin diagnostic drops below $1.1$ at a lower number of total samples), but in either case the diagnostic clearly indicates sufficient mixing between chains to satisfy the necessary invariance properties to anticipate that the posterior distributions are converging.

\subsection{Summary of posterior distribution}
\begin{figure}
    \centering
    \includegraphics[width=.5\textwidth]{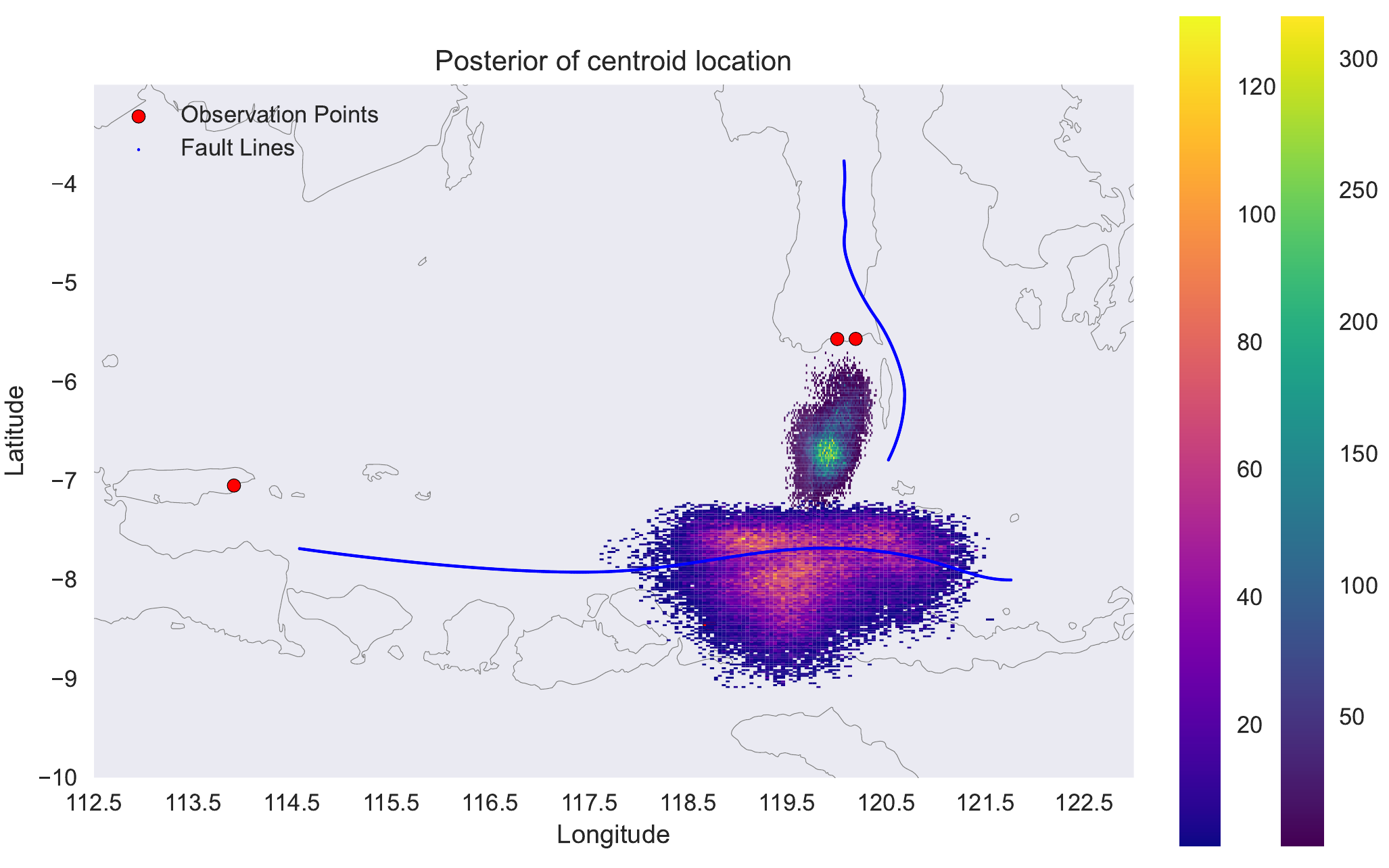}
    \caption{The posterior distribution of the earthquake centroid locations for both faults.  Note that those earthquakes originating on the Walanae/Selayar fault will be oriented primarily north-south (in line with the fault line itself) whereas those earthquakes along the Flores thrust will be oriented primarily east-west.  The prior distribution for the centroid of both of these faults is centered at 20km deep on the interior of the fault, i.e. to the west of the Walanae/Selayar fault line (blue curve) and south of the Flores fault line.}
    \label{fig:posterior_lat_lon}
\end{figure}

\begin{figure}
    \centering
    \includegraphics[width=.5\textwidth]{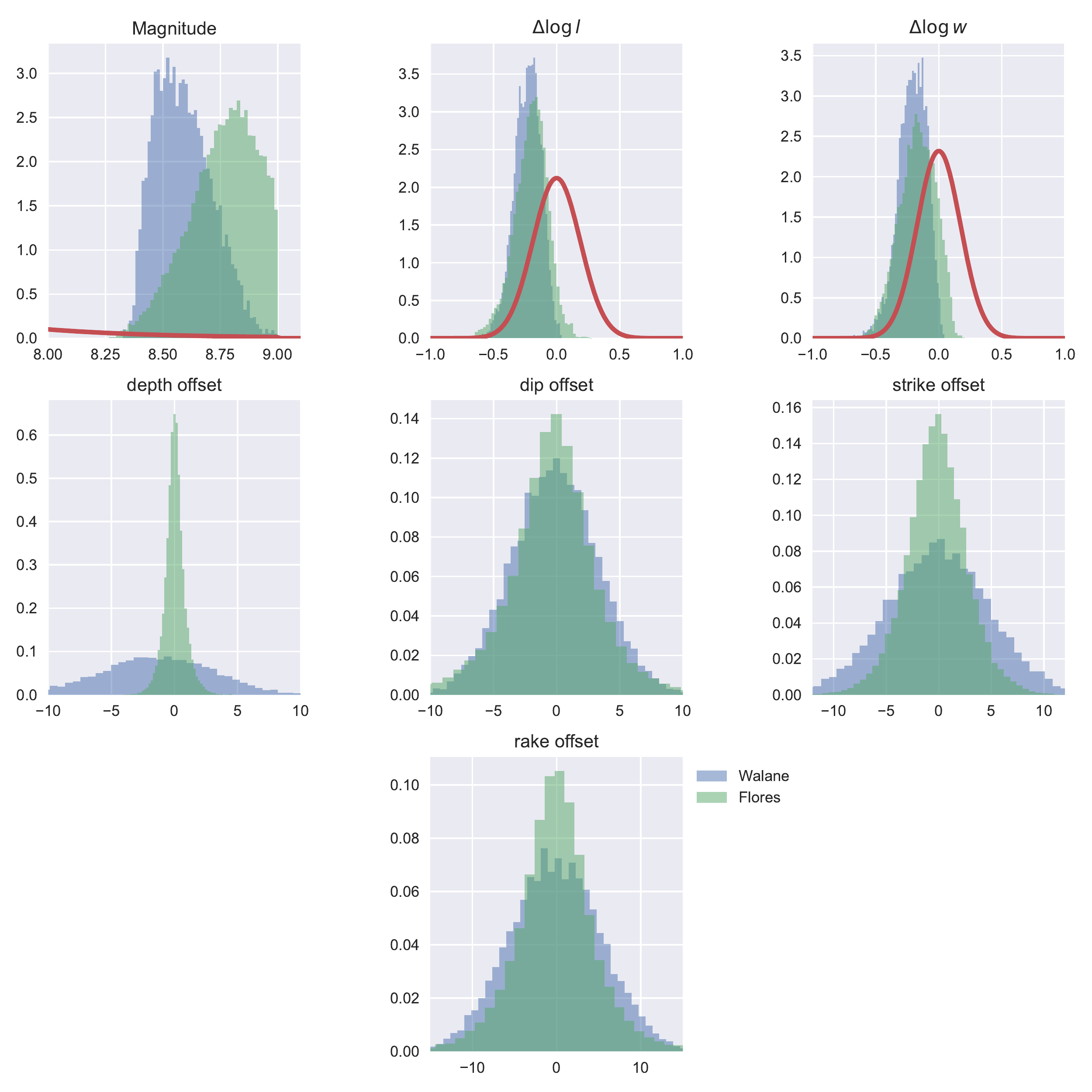}
    \caption{The posterior samples for all of the relevant sample parameters from both faults compared against the prior distributions.  We do not reproduce the prior distribution for the four offset parameters because the Flores prior for the offset parameters is centroid location dependent, i.e. the Gaussian process which models the Flores thrust yields an estimate of uncertainty at every point along the fault which is used to weight the prior distribution on the offset parameters for the Flores model.  This means that we can \emph{NOT} represent the prior distribution for the Flores fault along the offset parameters as a one-dimensional distribution without integrating against the position (a very costly exercise).}
    \label{fig:posterior}
\end{figure}

The primary description of the desired posterior distribution can be visualized via Figures \ref{fig:posterior_lat_lon} and \ref{fig:posterior}.  In particular, Figure \ref{fig:posterior_lat_lon} displays a histogram of the sampled centroids for both faults with the left colorbar representing the density of samples on the Flores thrust and the right colorbar representing the density of samples on Walanae/Selayar.  There are several items to note from this Figure alone:
\begin{itemize}
    \item The centroid location along the Walanae/Selayar fault is in a very concentrated location near $120^\circ$ longitude and $-6.5^\circ$ latitude.  In contrast the sampling along the Flores thrust is far less focused, with preferred centroid locations spanning a wide range of longitudinal values, and a relatively wide range of latitudes near $119.5^\circ-120^\circ$ longitude.
    \item The prior distribution on centroid location for the Flores thrust did not force the earthquake centroid to be on the `correct' side of the fault line (south of the blue curve in Fig. \ref{fig:posterior_lat_lon}).  This allowed for a surprising number of earthquake samples that were on the physically infeasible side of the fault (north of the blue curve), a region that appeared to actually be preferred to some extent by the sampling strategy employed here (there is a high concentration of centroids north of the blue curve in Fig. \ref{fig:posterior_lat_lon}).  This may indicate either that the Gaussian process prior is not sufficiently restrictive or (as discussed further below) the observational data prefers earthquakes centered north of the actual Flores fault.
    \item The centroids for the posterior on the Walanae/Selayar fault are on the `correct' side of the fault, but they are further south than the prior prefers, indicating that observations are better matched with an earthquake centroid further south than the modeled Walanae/Selayar fault extends.
    \item In addition, the most preferred centroid locations for the Flores thrust (at least those that lie on the `correct' side of the fault itself) line up with the curvature of the Walanae/Selayar fault.  That is, the centroid locations from the two faults nearly line up in a north-south line as if the Walanae/Selayar fault extended all the way to the Flores thrust.
\end{itemize}

Figure \ref{fig:posterior} depicts histograms of the sampled posterior distribution for all of the other sample parameters (omitting latitude and longitude which are depicted in Fig. \ref{fig:posterior_lat_lon}). We first note that the prior and posterior distributions for all four offset parameters are nearly identical for the Walanae/Selayar fault (even though the prior distribution is \emph{NOT} shown here).  The prior distribution for the Flores thrust is not independent in each of these parameters, as the sampled values are multiplied by the variance of the Gaussian process at each centroid location, and hence plotting a one-dimensional pdf of the prior would require integrating the full prior against the centroid position which is computationally prohibitive for visualization purposes alone.  For this reason we are unable to draw definitive conclusions about the influence of the observed data on the geometry of the Flores fault, however it is clear that the geometry of the Walanae/Selayar fault is not constrained by the data, i.e. the posterior simply recreates the prior distribution for these parameters.  Although we are unable to form the same comparison for the Flores fault we anticipate a similar result.  The rub of the matter is, our limited tsunami observations are not sufficient to constrain the geometry of the fault.

\begin{figure}
    \centering
    \includegraphics[width=.5\textwidth]{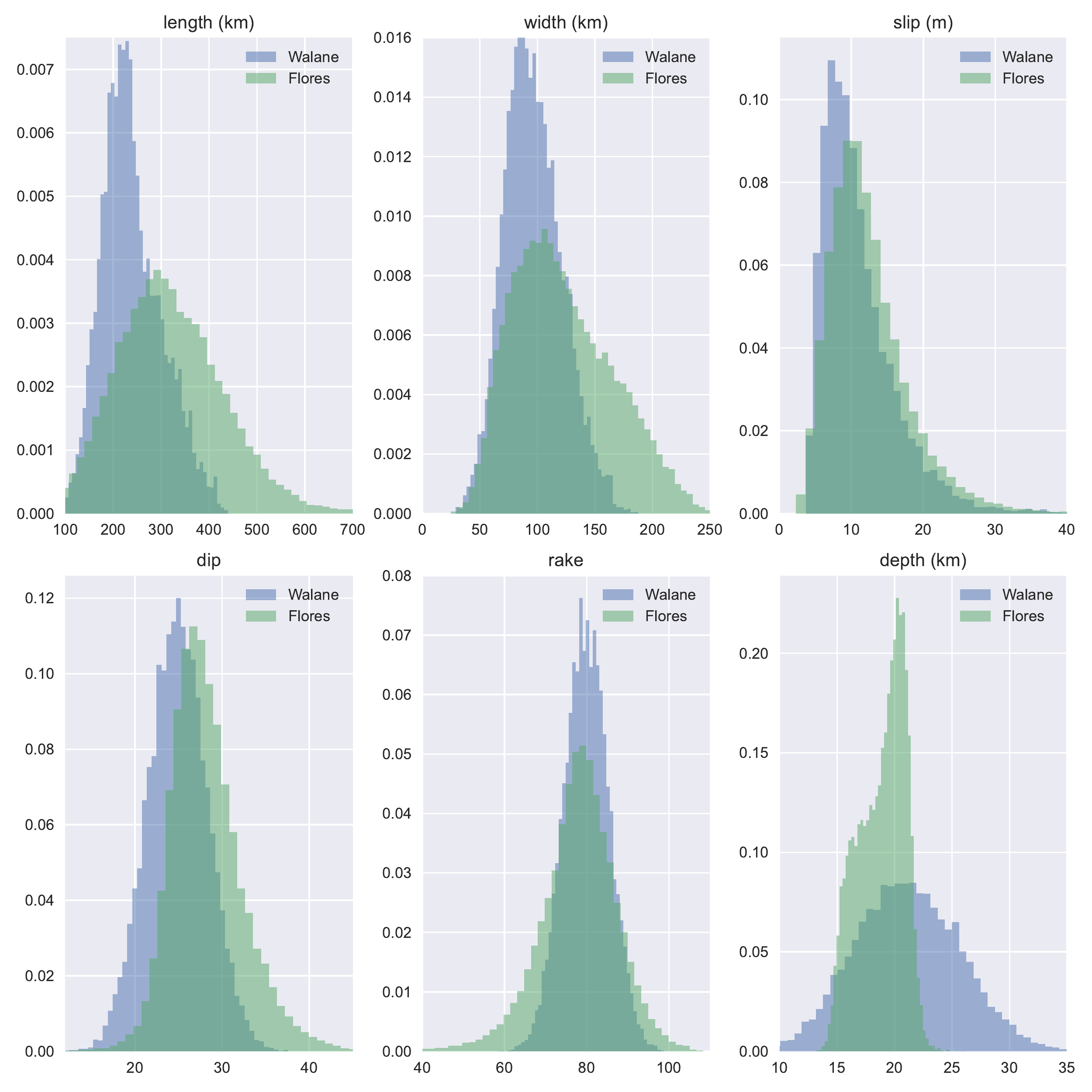}
    \caption{Histograms of the posterior distribution on the actual earthquake parameters (rather than sample parameters) for each fault.  The strike is omitted as it is fundamentally different for each fault (Walanae/Selayar runs north-south while Flores runs primarily east-west), and a comparison between the strike for the two posterior distributions is not informative.}
    \label{fig:EQ_params}
\end{figure}

On the other hand, there is a clear signal in both $\Delta \log l$ and $\Delta \log w$ that indicates that the observational data is a better match for smaller values of both of these parameters.  Smaller values of these parameters for a fixed magnitude corresponds to a larger slip length than expected, i.e. this indicates that the earthquakes that best match the data have very large slip as seen in Fig. \ref{fig:EQ_params}. We see that the most probable slip that matched the data for both faults was over $10m$ with a definite preference for larger slip.  The slip on the Walanae/Selayar posterior is slightly smaller, with a maximum probability estimate close to $8m$ rather than $10m$, and a slightly less positive bias toward larger slip, i.e. the Walanae/Selayar posterior is slightly more seismically sound.  This tendency toward an unexpectedly large slip was noticed in \cite{ringer2021methodological} for the 1852 Banda Sea earthquake in Eastern Indonesia where the Bayesian technique employed here was first introduced.  Future studies will consider the potential discretization effects and selection of hyper-parameters in the forward model that could lead to a preference for smaller rectangular area, large slip ruptures.

Fig. \ref{fig:EQ_params} also displays the other earthquake parameters derived from the posterior distribution.  Note in particular that a hypothesized Flores earthquake is less constrained in size as the length and width have a significantly wider histogram that extends to much larger values than earthquakes hypothesized for the Walanae/Selayar fault.  This result is likely because, as discussed in more detail below, the Flores posterior tends to favor extremely high magnitude earthquakes.  It is also interesting to note that in contrast to the magnitude derived parameters, the depth of the Flores posterior is more constrained than the depth for Walanae/Selayar.  This is likely a result of a more data-driven prior distribution on the Flores fault whereas the Walanae/Selayar prior is a nearly linear fit  and hence the depth values of the modeled fault are highly suspect.

This brings us to the final comparison between the two posterior distributions.  As previously described, the prior distribution on magnitude was the exponential Richter distribution that exponentially decays with growing magnitude as indicated by the red curve in the upper left plot of Fig. \ref{fig:posterior}.  Due to the size of the Flores and Walanae/Selayar faults, we also truncated the magnitude at $9.0$Mw to ensure physically reasonable earthquakes were observed.  As shown the observational accounts best matched with earthquakes of extremely large magnitude, particularly along the Flores thrust where the most probable magnitude is near $8.8$Mw, with a clear preference toward the cutoff magnitude of $9.0$Mw.  In contrast, although the earthquakes sampled from the Walanae/Selayar posterior were also quite large for the size of the Walanae/Selayar fault with the most likely value near $8.5$Mw, the Walanae/Selayar posterior did not have as much of a positive bias toward extremely high magnitude events.  In fact, the Walanae/Selayar posterior preferred earthquakes around $8.5$Mw, which although large for the fault in question, is far more likely than an $8.8$Mw event.

The high magnitude preference for both posterior distributions is in line with the observation made previously that the slip was quite large for both of these earthquakes.  Neither the Walanae/Selayar fault nor the Flores thrust are large enough to sustain an $8.5$Mw earthquake with the standard relationship maintained between length, width and slip.  However the observational data indicates that large magnitude events are necessary for the tsunami observations to match.  The apparent trade off here is satisfied with large (but not extreme) magnitude earthquakes that are shorter and narrower, but with very high average slip length (recall that our model of slip requires an instantaneous, uniform slip distribution across the entire rupture). An alternative hypothesis is that the earthquake was triggered on both the Walanae/Selayar and Flores faults, perhaps allowing for a smaller magnitude event on both nearly simultaneously.

\subsection{Comparison of posterior predictive to observation probabilities}
Figure \ref{fig:Observations} depicts the histograms of the posterior predictive (output of each simulation for the observational data points) relative to the original observational probabilities.  Each of the six observations have particular characteristics that are of interest in this setting:
\begin{figure}
    \centering
    \includegraphics[width=.5\textwidth]{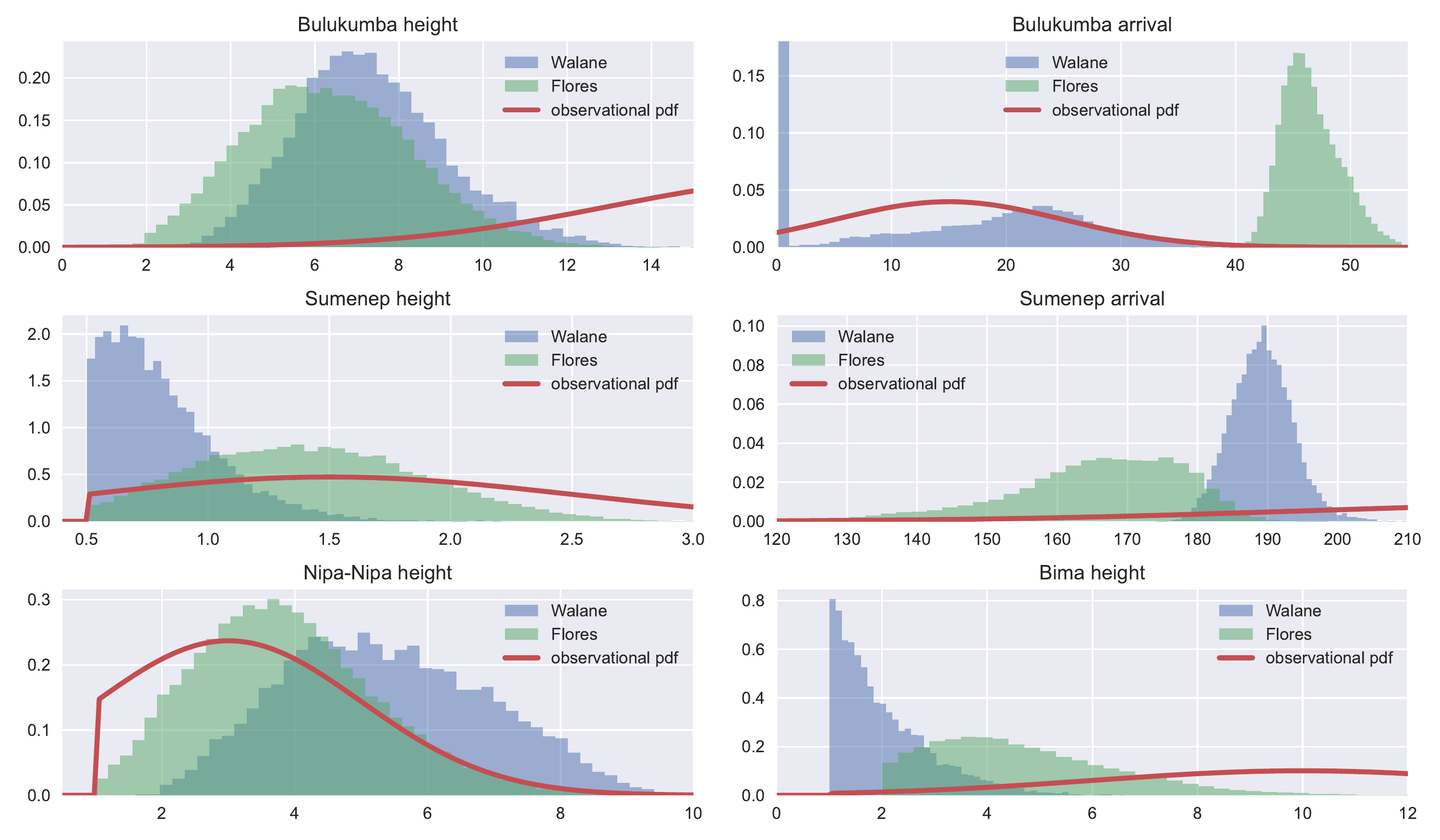}
    \caption{Posterior samples from both hypothesized faults plotted against the observational probabilities assigned to each observation.}
    \label{fig:Observations}
\end{figure}
\begin{itemize}
    \item The extreme wave height from the observation in Bulukumba is clearly not achieved for either posterior distribution, leading us to believe that either the historical observation which claimed a wave height of 60-80 feet (18-24 meters) was over-exaggerated, or some other nonlinear, local effects were at play.  In particular, it is possible that a submarine landslide caused by the earthquake could generate a wave of this magnitude at least locally.  This hypothesis is reasonable when we consider that the wave heights at Nipa-Nipa generated from either fault are near the observational probabilities, noting that Nipa-Nipa is geographically very close to Bulukumba so that it is highly unlikely that Bulukumba would have a wave near 20 m whereas Nipa-Nipa only sustained one of 4-5 m.
    \item The arrival time in Bulukumba has a few peculiarities.  The large number of arrival times at time $0$ for the Walanae/Selayar posterior arise because a significant number of the Walanae/Selayar earthquakes have a rupture zone overlapping the observation locations at Bulukumba so that the sea surface is instantaneously shifted i.e. the wave seemingly arrives instantaneously although this arrival time isn't the actual wave arriving, but just the initial disturbance from the earthquake.  Beside these events, it is apparent that the posterior predictive from the Walanae/Selayar fault matches the observational distribution quite well for the Bulukumba arrival time whereas the Flores posterior indicates a much longer arrival time to Bulukumba than anticipated.
    \item The observational distribution for wave height at Sumenep appears to better match the Flores posterior predictive except that for this particular observation we must recall that the only statement was that the wave was observed at Sumenep, i.e. the observational distribution at this location is not very precise.
    \item The arrival time at Sumenep clearly doesn't agree well with either posterior predictive, but it is also certain that the Walanae/Selayar posterior is a much better fit than the Flores as waves originating from the Walanae/Selayar fault take over 3 hours to arrive at Sumenep while most tsunamis originating from Flores arrive just under 3 hours.  This particular comparison should not be weighted too heavily though, as neither fault generates a tsunami whose initial wave arrives in 4 hours which is the observational value only after assuming that Eastern Java and Sulawesi are in two different time zones.  A partial explanation for this is that the initial wave is not the one recorded in the historical record, but that a secondary wave is the one observed in Sumenep.   We did not collect the arrival times for the secondary waves for all $200,000+$ simulated earthquakes, but from repeat simulation of a few events we did note that some of the later waves from both the Walanae/Selayar and Flores faults were larger than the initial wave with corresponding arrival times exceeding 210 minutes (for Flores) and 240 minutes (for Walanae/Selayar). Low resolution bathymetry may also play a role in faster predicted versus actual arrival times.
    \item The posterior predictive wave height at Bima is also not a great match with the observational data for either fault, although earthquakes generated from the Flores thrust match the observation better.  In essence the generated earthquakes are under-estimating the wave height in Bima.  There are several potential reasons for this, one of which may simply be that the bathymetric resolution isn't sufficient to capture the amplification of the wave entering the bay.  Even with the type of amplification that may occur, the Walanae/Selayar posterior clearly underestimates the wave height at Bima as it is hard to imagine a $2m$ wave having sufficient buoyancy and force to `fling' ships far inland.
\end{itemize}

\subsubsection*{Which fault fits the observations better?}
The discussion above gives credence from some observations for each of the potential posteriors, i.e. some parts of the posterior predictive are supportive of the Walanae/Selayar hypothesis and some support the Flores Thrust.  To make a quantifiable comparison between the two hypotheses, we propose a \emph{novel} approach using the dual posterior predictive combined with the observational probabilities to obtain a relative probability of which fault best matches the observational data.
\begin{enumerate}
    \item Using all of the data from the two posterior predictives, we trained a binary classifier whose input was the wave heights and arrival times for all 6 observed data points.  The data was divided into two distinct classes defined by the fault the earthquake originated from.
    \item Once we trained a sufficiently accurate classifier, we drew samples from the observational probabilities (as shown in red in Fig \ref{fig:Observations}) and use the previously trained classifer to determine which fault these observations were most likely generated from.
\end{enumerate}

To train the binary classifier, we first took all samples from both faults, removed those that did \emph{not} have finite log posterior, and randomly selected $70\%$ of the remaining samples for the training set and the remaining $30\%$ for the testing set to give:
\begin{itemize}
    \item 162,862 samples for the training set: 89,554 originating from the Flores fault and 73,308 from Walanae/Selayar.
    \item 69,798 samples for the testing set: 38,136 from the Flores fault and 31,662 from Walanae/Selayar.
\end{itemize}
Both the training and testing sets were normalized using sklearn's preprocessing package to reduce the amount of bias in the fit. Then rather than rely on a single classifier, we trained two different classifiers of different architectures to ensure the results were consistent:
\begin{itemize}
    \item We trained a binary logistic classifier via XGBoost (with all default settings).  This classified all but 5 of the samples from the test set correctly for an accuracy of 99.993\%.
    \item We also trained a random forest classifier again with all default settings from sklearn, which only misclassified 3 samples from the test set for an accuracy of 99.996\%.
\end{itemize}

To identify which fault best matched the source we sampled 1 million data points from the observational probabilities depicted in Fig. \ref{fig:Observations}.  After normalization, each set of these data points was then fed through the classifier and a fault source was prescribed according to the classifier.  The random forest classifier selected Walanae/Selayar as the source 94.4\% of the time and the logistic regression classifier selected Walanae/Selayar 98.4\% of the time.

These probabilities should be considered in the proper context however.  We can summarize the above results in the following probabilistic statement:
\begin{quote}
    Given the hypothesis that either the Walanae/Selayar or Flores faults were the source of the 1820 tsunami, and provided that the observational probabilities depicted in Fig. \ref{fig:Observations} are realistic, the random forest classifier is 94\% confident that the Walanae/Selayar fault was the source.
\end{quote}
This does not say that the authors are 94\% confident that the Walanae/Selayar fault is the source, but that if the only hypothesis considered is that one of the two faults separately generated the tsunami, then we are.  In other words, if we are certain that one of these two faults generated the observations, we are quite certain it had to originate from the Walanae/Selayar fault and \emph{NOT} the Flores thrust.  The rub of the matter is that this yields a conditional probability, i.e. we have a roughly 94\% probability that Walanae/Selayar was the source of the tsunami, given that the source was either Walanae/Selayar or Flores.  Even so, as noted above in our explicit discussion on different parts of the posterior predictive, this is not a completely satisfactory hypothesis.

Careful investigation of either classifier described above implies that the wave arrival time in Bulukumba seems to be the determining factor.  While all the other observables may be mildly more or less probable to identify with either fault the Bulukumba arrival time histograms for the Walanae/Selayar and Flores faults are clearly separated with the Walanae/Selayar data matching the observational distribution much better.

\subsubsection*{So who is at fault?}
In summary:
\begin{itemize}
    \item Neither posterior appears to match the wave height at Bulukumba, which we anticipate is a result of an over-inflated observation or an additional local source.
    \item Neither fault appears to match the arrival time at Sumenep although Walanae/Selayar certainly is closer than Flores.
    \item Neither fault matches the wave height at Bima well, but Flores is much closer than Walanae/Selayar.
    \item Finally the arrival time at Bulukumba is clearly better fit with an earthquake originating from the Walanae/Selayar fault.
\end{itemize}
     Hence, from purely viewing the posterior predictive, the selection of which fault was the most likely source of the tsunami, is uncertain at best.  In overall probability (likelihood) it is apparent that the Walanae/Selayar fault is a better fit to the assigned observational distributions, but claiming the tsunami was generated on Walanae/Selayar would severely discount the detailed observation at Bima.  On the other hand, a tsunami generated on Flores clearly misses the wave arrival time in Bulukumba and has a significantly poorer fit to the arrival time in Sumenep.

As a final note, we recognize that the Walanae/Selayar fault is a more likely candidate to match the shaking intensity recorded at Bulukumba, where as noted above, the canons were said to jump from their mounting.  This implies a peak ground acceleration exceeding 1 G at a minimum, which would be highly unlikely for an earthquake originating over 200km away (the distance from Bulukkumba to the Flores thrust).  Hence, although we have not made use of the shaking observations in our computations, our final result that indicates the earthquake was more likely generated along the Walanae/Selayar fault, is consistent with those observations.

\section{Discussion and Conclusion}\label{sec:conclusions}
The Bayesian approach toward identifying characteristics of earthquakes using anecdotal historical accounts of tsunamis first introduced in \cite{ringer2021methodological} has been applied to the 1820 south Sulawesi earthquake and consequent tsunami.  Using the Bayesian framework, we have simulated close to 200,000 different events, searching through parameter space for earthquakes that probabilistically best match the interpreted historical record.  Hypothesizing that the earthquake originated purely from either the Flores or Walanae/Selayar faults does not yield a posterior distribution that appears to match the data perfectly, although we have strong statistical reasoning to assert that the Walanae/Selayar fault was far more likely to be the source than the Flores thrust.

To further investigate this event and hopefully ascertain the source of the recorded tsunami, we will next consider two potential hypotheses:
\begin{itemize}
    \item A landslide near where the Walanae/Selayar Fault goes offshore (likely very near Bulukumba itself) that can produce the significant wave heights recorded near the Fort.
    \item The dual rupture of both the Walanae/Selayar and Flores faults.  Although a time-dependent rupture is clearly more physically relevant, we will restrict our attention to an instantaneous rupture of both faults simultaneously, as it is unlikely that our limited observations of tsunami impacts will provide enough detail to constrain a more sophisticated rupture model.
\end{itemize}
Further extensions of this Bayesian approach to historical tsunamis will be carried out both to improve the sampling procedure, and to investigate other events with the goal of providing a thorough description of the past seismicity in the Indonesian region.

\section*{Acknowledgments}
Beyond the authors listed on this manuscript, there are several students who have come through BYU that have contributed to the formation of these results.  These include but are not limited to: H. Ringer, S. Giddens, M. Morrise, C. Noorda, J. Callahan, K. Lightheart, C. Kesler, and C. Herrera. In particular, we wish to thank N. Thompson for his assistance with improving the presentation of some of the figures reported here.  We also would like to thank N. Glatt-Holtz for inspiring this work several years ago, and useful conversations with R. LeVeque, V. Martinez, C. Mondaini, A. Holbrook, as well as a host of other colleagues who have provided useful feedback and guidance over the years.

We also acknowledge generous financial support from the BYU Mathematics and Geology Departments as well as the College of Physical and Mathematical Sciences. JAK would like to acknowledge the support of the National Science Foundation under grant DMS-2108791.
%\end{acknowledgments}

\section*{Data Availability}
All of the relevant code that generated the data provided in this article appears in \cite{zenodov1_1}.  The data itself can be viewed via the graphical interface provided at \url{http://tsunami.byu.edu/whitehead-lab}.
%\end{dataavailability}

\bibliographystyle{gji}
\bibliography{tsunami.bib}

\end{document}